\documentclass[sigconf, screen]{acmart}

\AtBeginDocument{
  }

\renewcommand\footnotetextcopyrightpermission[1]{}

\copyrightyear{none}
\acmYear{none}
\acmDOI{none}

\usepackage{amsfonts}
\usepackage{nicefrac}
\usepackage{microtype}
\usepackage{xcolor}
\usepackage{pifont}
\usepackage{colortbl}
\usepackage{multirow, makecell}

\usepackage{subcaption}
\usepackage[most]{tcolorbox}
\usepackage{listings}

\author{Chaojie Sun}
\affiliation{%
  \institution{\mbox{Zhejiang University of Technology}}
  \city{Hangzhou}
  \country{China}
}
\email{sunchaojie@zjut.edu.cn}

\author{Bin Cao}
\affiliation{%
  \institution{\mbox{Zhejiang University of Technology}}
  \city{Hangzhou}
  \country{China}
}
\email{bincao@zjut.edu.cn}

\author{Tiantian Li}
\affiliation{%
  \institution{\mbox{Zhejiang University of Technology}}
  \city{Hangzhou}
  \country{China}
}
\email{ttli89@zjut.edu.cn}

\author{Chenyu Hou}
\affiliation{%
  \institution{\mbox{Zhejiang University of Technology}}
  \city{Hangzhou}
  \country{China}
}
\email{houcy@zjut.edu.cn}

\author{Ruizhe Li}
\affiliation{%
  \institution{\mbox{University of Aberdeen}}
  \city{Aberdeen}
  \country{UK}
}
\email{ruizhe.li@abdn.ac.uk}

\author{Jing Fan}
\affiliation{%
  \institution{\mbox{Zhejiang University of Technology}}
  \city{Hangzhou}
  \country{China}
}
\email{fanjing@zjut.edu.cn}

\begin{document}

\title{FGTR: Fine-Grained Multi-Table Retrieval via Hierarchical LLM Reasoning}

\begin{abstract}
With the rapid advancement of large language models (LLMs), growing efforts have been made on LLM-based table retrieval. However, existing studies typically focus on single-table query, and implement it by similarity matching after encoding the entire table. These methods usually result in low accuracy due to their coarse-grained encoding which incorporates much query-irrelated data, and are also inefficient when dealing with large tables, failing to fully utilize the reasoning capabilities of LLM. Further, multi-table query is under-explored in retrieval tasks. To this end, we propose a hierarchical multi-table query method based on LLM: \underline{\textbf{F}}ine-\underline{\textbf{G}}rained Multi-\underline{\textbf{T}}able \underline{\textbf{R}}etrieval (\textbf{FGTR}), a new retrieval paradigm that employs a human-like reasoning strategy. Through hierarchical reasoning, FGTR first identifies relevant schema elements and then retrieves the corresponding cell contents, ultimately constructing a concise and accurate sub-table that aligns with the given query. To comprehensively evaluate the performance of FGTR, we construct two new benchmark datasets based on Spider and BIRD . Experimental results show that FGTR outperforms previous state-of-the-art methods, improving the $F_2$ metric by 18\% on Spider and 21\% on BIRD, demonstrating its effectiveness in enhancing fine-grained retrieval and its potential to improve end-to-end performance on table-based downstream tasks.
\end{abstract}




\maketitle
\makeatletter
\fancyhead{}

\section{Introduction}
Table reasoning tasks involve performing reasoning and deriving answers over structured tables, widely used in healthcare, finance, and corporate financial reporting. Distinct from textual data, tabular data exhibits a two-dimensional structure with complex logical relationships among rows, columns, and cell contents, posing significant NLP challenges. Common tasks include table question answering \citep{herzig2020tapas,liu2021tapex}, table fact verification \citep{shi2021logic,gu2022pasta,zhou2022table}, table-to-text generation \citep{bian2023helm}, and table semantic parsing \citep{dong2023c3}. As retrieved tables serve as critical inputs for reasoning, effective table information retrieval becomes pivotal \citep{chen2024table}. Reducing redundant table content directly enhances model accuracy, computational efficiency, and complex reasoning capability.

In natural language processing (NLP), LLMs have made significant strides, opening new frontiers for table reasoning tasks. However, their application to structured data still faces persistent challenges due to inherent limitations. First, LLMs’ inherent context window constraints pose critical limitations when processing large-scale tables containing thousands of rows or columns despite their proficiency in conventional text tasks. Common mitigation strategies for large tables include table segmentation or setting a token budget for content extraction \citep{sui2023tap4llm,chen2024tablerag}, which risks information loss and subsequent degradation of reasoning accuracy. Second, LLMs demonstrate constrained capabilities in comprehending tabular structures and logical relationships\citep{fang2024large}, attributed to:

\begin{figure}[t!]
  \centering
  \begin{subfigure}[t]{0.55\linewidth}
    \centering
    \includegraphics[width=\linewidth]{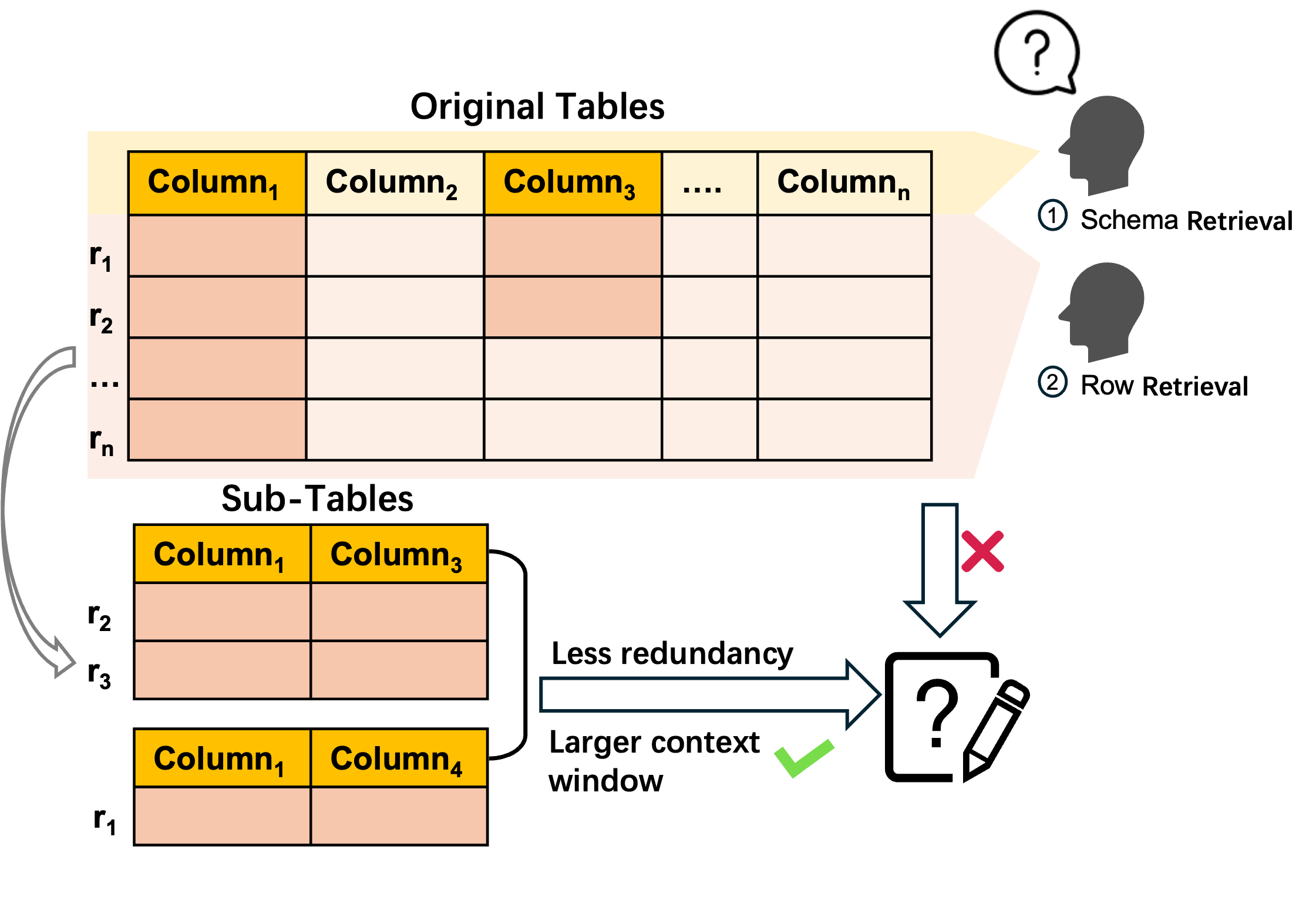}
    \caption{}\label{subfig:a}
  \end{subfigure}\hfill
  \begin{subfigure}[t]{0.45\linewidth}
    \centering
    \includegraphics[width=\linewidth]{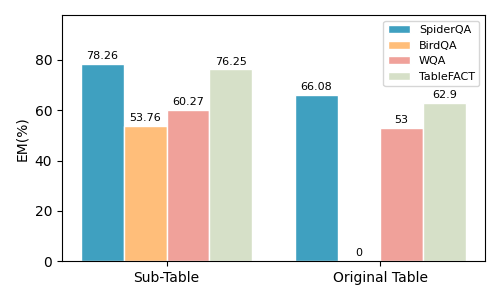}
    \caption{}\label{subfig:b}
  \end{subfigure}
  \caption{Illustration of the human cognitive process for table comprehension and the motivation for fine-grained retrieval. (a) Humans typically adopt a hierarchical approach: they first scan schemas to identify relevant columns and then locate the specific rows needed, effectively creating a mental sub-table. (b) Performance comparison on BirdQA demonstrating the necessity of retrieval. Feeding the compact, retrieved sub-table significantly boosts QA accuracy, whereas feeding the full table leads to context window overflow and catastrophic failure.}\label{fig:two-side}
\end{figure}

\noindent
\textbf{Prior Knowledge Deficiency.} While existing LLMs predominantly employ autoregressive mechanisms\citep{fang2024large} trained on sequential textual data to acquire human-like linguistic understanding, Serializing two-dimensional tables into linear text sequences compromises their structural integrity. This conversion process creates excessive reasoning spans between originally vertically adjacent cells, impeding cross-cell information alignment and structural comprehension.

\noindent
\textbf{Structural Disorder.} Unlike naturally structured modalities (e.g., text with syntactic patterns), tabular data as semi-structured information lacks unified formatting standards and exhibits ambiguous inter-schema relationships. The absence of inherent structural regularities deprives LLMs of foundational priors for interpreting feature correlations, contrasting sharply with their proficiency in processing naturally ordered data types.

Hence, it is necessary to overcome these issues through fine-grained multi-table retrieval, as table-based reasoning constitutes a complex task requiring human-like cognitive processes and fine-grained comprehension\citep{ye2023large}. However, existing studies still exhibit two main limitations. First, current approaches primarily encode the entire table\citep{chen2024tablerag} or encode rows and columns\citep{lin2023inner,sui2023tap4llm} to retrieve question-relevant contents and construct sub-tables. Such methods depend heavily on holistic table representations, resulting in limited understanding of table structures and low encoding efficiency. In contrast, as illustrated in Figure~\ref{fig:two-side}, humans typically first attend to the table’s schema structure—including column names, attributes, and sample entries—to rapidly grasp the informational framework and overall theme. On this basis, they subsequently focus on the columns required by the question, locate specific cell contents, and obtain the relevant rows needed to solve the problem. Second, existing fine-grained table retrieval methods tend to oversimplify real-world scenarios by restricting retrieval to a single table, neglecting the fact that in practical cases, the required information is often distributed across multiple tables and must be integrated through inter-table joins and relationships \citep{chen2024table}.

In this paper, we propose a novel \textit{Fine-Grained Multi-Table Retrieval}(FGTR), a framework leveraging LLMs’ reasoning capabilities to hierarchically execute schema retrieval followed by cell retrieval for precise information localization. FGTR is designed to simulate the human cognitive process of reading tables: it first conducts schema retrieval to identify columns relevant to a given query, and then performs cell retrieval to further locate the specific information required. This decompositional reasoning approach effectively transfers the textual reasoning strengths of LLMs to the domain of complex two-dimensional tabular data, allowing for structured and interpretable table reading while overcoming the model’s inherent issues of disorder and lack of prior structural knowledge when understanding tables. Furthermore, our framework extends its capabilities to complex multi-table scenarios by optimizing the retrieval of crucial joining keys between tables, enabling robust cross-table reasoning.

Our contributions can be summarized as follows:

\begin{itemize}
  \item We developed FGTR, a human-like hierarchical retrieval strategy for fine-grained table retrieval, which demonstrates superior retrieval performance on real-world multi-table datasets while consuming fewer tokens. Moreover, FGTR applies to any table processing task framework. We conducted comprehensive downstream table task experiments and ablation studies to validate the effectiveness of each component of FGTR.
  \item Based on the Spider\citep{yu2018spider} and BIRD\citep{li2024can} datasets, we built two new benchmark datasets for fine-grained table retrieval, aiming to standardize the evaluation of accuracy and recall in fine-grained table retrieval tasks.
  \item We also explored a comparison between open-source and closed-source large models in the context of table retrieval strategy performance and performed knowledge distillation on open-source models to improve the retrieval performance of smaller models.
\end{itemize}

\section{Related Work}
LLMs have recently been widely adopted in table-related reasoning tasks such as TableQA\citep{chen2022large,zhao2024tapera,yu2025table}. However, due to limited context windows and the inherent difficulty of reasoning over large structured tables\citep{liu2023lost}, fine-grained table retrieval becomes essential for efficient TableQA. Compared to open-domain table retrieval\citep{herzig2021open,chen2024table,christensen2025fantastic,zou2025gtr} which targets entire tables,fine-grained table retrieval Focuses on identifying specific schema elements or cell contents that are most relevant to the query. It serves as a critical preliminary step for large-scale TableQA, aiming to provide a concise and relevant context for the downstream LLM. Existing work is primarily divided into two main categories:

\noindent
\textbf{Schema-based Retrieval.} This approach primarily serves Text-to-SQL tasks\citep{dong2023c3,talaei2024chess,cao2024rsl,shkapenyuk2025automatic}. Since the core of SQL generation lies in understanding the table's structure, this type of work retrieves relevant schema subsets by matching the query with the table's metadata. However, these methods typically do not consider the specific values within cells, making them unsuitable for question-answering tasks that require content reasoning.

\noindent
\textbf{Cell-based Retrieval.} This approach aims to directly locate cells relevant to the query to construct a smaller, more information-dense sub-table. Dater\citep{ye2023large}, Binder\citep{cheng2022binding} and PieTa\citep{lee2024piece} utilize large models to retrieve query-relevant content, but these methods fail to fundamentally circumvent the inherent challenges LLMs face when processing long tables. CLTR\citep{pan-etal-2021-cltr},ITR\citep{lin2023inner} and Tab4LLM\citep{sui2023tap4llm} encode the rows and columns of a table to determine their relevance to the query, thereby forming a sub-table. Tabsqlify\citep{nahid2024tabsqlify} leverages SQL queries to retrieve sub-tables. TableRAG\citep{chen2024tablerag} synchronously retrieves and synthesizes a sub-table by encoding cell values and the schema. The performance of these methods is limited by semantic matching, making it difficult to effectively handle queries that require multi-step reasoning or complex logical associations. Additionally, encoding the entire table incurs high computational overhead.

Previous work on fine-grained table retrieval has predominantly focused on single-table settings. Our proposed FGTR framework is the first to extend this problem to the more challenging multi-table question answering (Multi-table QA) scenario. Furthermore, our hierarchical reasoning strategy guides the LLMs to perform logical inference, progressively narrowing the retrieval scope and avoiding an exhaustive scan of the entire table's content. This approach not only enhances retrieval efficiency but also enables our model to achieve superior performance over existing methods in complex cross-table retrieval tasks.

\begin{figure*}[t!]
  \centering
  \includegraphics[width=0.85\textwidth]{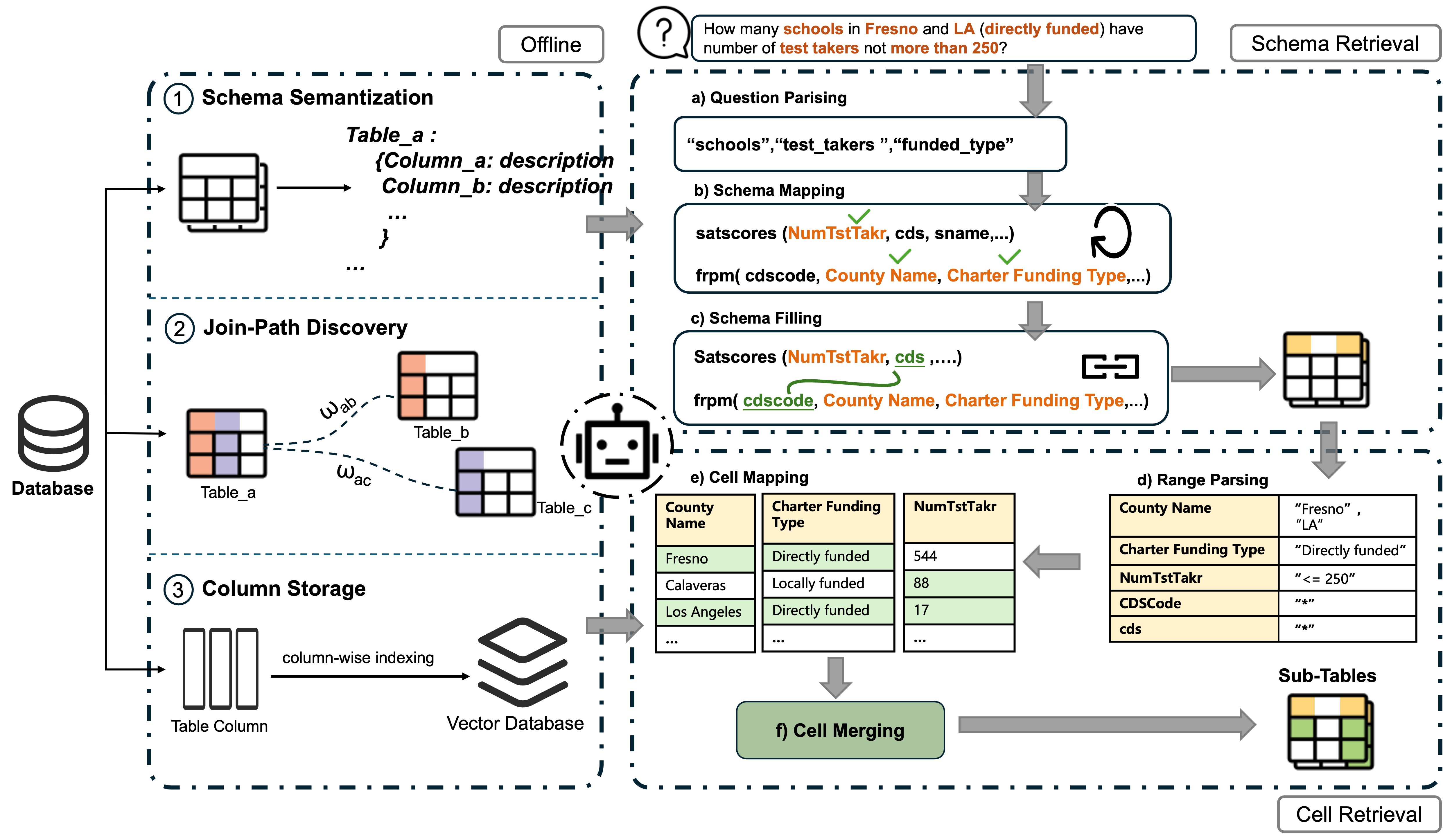}
  \caption{The overall architecture of our FGTR framework, which consists of an offline preprocessing phase and a two-stage online retrieval process: Schema Retrieval and Cell Retrieval.}
  \label{fig:struct}
\end{figure*}

\section{Problem Formulation}
A fine-grained table retrieval problem can be formally defined as follows. Given a database \( \mathcal{D} = \{T_1, T_2, \dots, T_d\} \) and a natural language query \( Q \), the objective is to retrieve a set of sub-tables \( \{T'_i\}_{i=1}^k \) (where \( k \leq d \)), with each sub-table \( T'_i \) containing the minimal yet sufficient information from its original table \( T_i \) to answer \( Q \). Let a table \( T_i \in \mathcal{D} \) be defined by a set of columns \( C_i = \{c_{i,1}, c_{i,2}, \dots, c_{i,n}\} \). The retrieval process is performed in two stages. First, in Schema Retrieval, for each table \( T_i \), a set of query-relevant columns is identified:
\begin{equation}
S_i = \{c_x \in C_i \mid \text{Rel}_c(c_x, Q) = \text{True} \},
\end{equation}
where \( \text{Rel}_c \) is a function determining the relevance of a column \( c \) to the query \( Q \). Second, in Cell Retrieval, for each table with a non-empty \( S_i \), a set of relevant rows is identified:
\begin{equation}
\mathcal{R}_i = \{r_{.,j} \mid \text{Rel}_r(c_{x,y}, Q, S_i) = \text{True} \},
\end{equation}
where \( \text{Rel}_r \) evaluates whether the \( j \)-th row, \( r_{i,j} \), contains information pertinent to \( Q \) given the selected columns \( S_i \). The final output is the set of all non-empty sub-tables \( \{ T'_i \} \), where each \( T'_i \) is formed by projecting the columns in \( S_i \) for all rows indexed by \( \mathcal{I}_i \) from the original table \( T_i \).

\section{Fine-Grained Table Retrieval}
Figure~\ref{fig:struct} illustrates the overall architecture of our proposed FGTR framework. The framework operates in two distinct phases. In the offline phase, FGTR processes the target database through three parallel steps: (1) Schema Semantization, (2) Join-Path Discovery, and (3) Columnar Storage construction, establishing a structured foundation independent of specific queries. Subsequently, in the online phase, given a user query, the system executes a hierarchical retrieval process based on the preprocessed database: first identifying relevant schemas and then retrieving the corresponding cells. This hierarchical workflow ultimately yields a fine-grained sub-table tailored to the specific query.
\subsection{Offline Preprocessing}
Our method involves two offline preprocessing steps before online inference.
First, we process the schema information of each table $T_i$ in the collection $D$, we utilize the metadata of each table $T_i$—including column names, data types, the top three most frequent example values, and the longest and shortest string examples—to prompt a LLMs to generate natural language descriptions for the table and each of its columns. The generated outputs are stored in JSON format as a structured and interpretable table schema (see Appendix~\ref{app:prompt} for details).

Second, to handle implicit joins in multi-table queries, we pre-compute potential foreign key relationships across the database. Since explicit foreign key constraints are often missing, we estimate the connection weight \( \omega_{ij}^{kl} \) for every pair of columns (\(c_k, c_l\)) between any two tables \(T_i\) and \(T_j\), following the method from \citep{chen2024table}:
\begin{equation}
\omega_{ij}^{kl} = \big( s(c_k, c_l) + j(c_k, c_l) \big) \times \max \big( u(c_k), u(c_l) \big),
\end{equation}
where \( s(c_k, c_l) \) represents the semantic similarity between two columns, covering both their names and fnatural language descriptions, calculated using OpenAI's text-embedding model, \( j(c_k, c_l) \) is their instance-level Jaccard similarity, and \( u(c_k) \) indicates value uniqueness, which is defined as the ratio between the number of unique values and the total number of rows.A higher weight suggests a stronger primary-foreign key link.

Third, to facilitate efficient cell retrieval, we preprocess and index all table cells into a vector database. For each non-numeric column, we aggregate cells with identical values and encode each unique cell value using a semantic encoder. Hierarchical Navigable Small World (HNSW) \citep{malkov2018efficient} is an effective approximate nearest neighbor search algorithm that can efficiently retrieve a large number of entries from large-scale tables, ensuring low-latency online retrieval. In our preprocessing stage, we utilize HNSW to build an index for the unique values within each column.

\subsection{Schema Retrieval}

The objective of schema retrieval is to extract query-relevant schemas \( S_i \) from tables \( T_i \). Rather than relying on inefficient full-database encoding, FGTR implements a decompositional reasoning framework. We introduce a “parsing–mapping–filling” strategy, inspired by Column Type Annotation (CTA) and schema linking. This approach first decomposes the query into semantic components and maps them to specific schemas, then enriches the set with relational keys. By explicitly incorporating these join paths, our strategy guarantees a comprehensive schema set, significantly enhancing retrieval accuracy for complex downstream reasoning.

\noindent
\textbf{Question Parsing.}
Breaking down complex problems into smaller, stepwise subproblems helps large models perform reasoning more efficiently. This approach is highly effective in numerical and commonsense reasoning tasks\cite{ye2023large,chen2024tablerag}. Suppose both the problem and table schema are directly provided to the LLM for reasoning. In that case, the inference process becomes overly complex, making distinguishing the key relevant information difficult. Therefore, we leverage the large model to extract the key elements from the user’s query and infer which schema is likely to contain the information required for those key elements, these elements typically include the keywords, keyphrases, and named entities in the question, providing an essential reference for subsequent reasoning steps. For instance, from the query \emph{“How many schools in Contra Costa and LA (directly funded) have number of test takers not more than 250?}”, the LLM would infer a potentially relevant schema, as illustrated in Figure~\ref{fig:struct} (a).

\noindent
\textbf{Schema Mapping.}
In the schema mapping stage, the key elements extracted during question parsing are logically aligned with the database schema through LLM-based reasoning. The core objective of this stage is to leverage the model’s reasoning capability to map the required columns identified in the parsing step onto the actual table schema, utilizing the processed schema information. It is important to note that while the parsing results serve as a critical reference for schema mapping, they are not the sole determinant. Schema mapping extends beyond simple one-to-one correspondence and may include one-to-many mappings. For example, a parsed element such as "name" may correspond to multiple columns like "first\_name" and "last\_name" in the actual table. In addition, the LLM may need to infer certain columns that are not explicitly identified during the parsing stage.

In this process, recall is prioritized over precision\citep{li2023resdsql}, ensuring that all potentially relevant information is retrieved for downstream reasoning tasks, even at the cost of redundancy. To achieve this, we instruct the LLM to generate K iterations of schema retrieval results, record the occurrence frequency of each column, and employ a voting mechanism to select the columns exceeding a consistency threshold $\theta$ as the final output. This approach mitigates random variability from single-pass inference while maximizing both recall and precision. Moreover, since the order of prompts can significantly influence LLM judgment \citep{liu2023lost}, we randomize the ordering of columns within each table and across multiple tables when constructing the prompts to reduce positional bias. The final output of this stage is the set of query-relevant columns \( S \), determined by:
\begin{equation}
S = \{c \in \bigcup_{i=1}^{d} C_i \mid \text{Freq}(c) \geq \theta \cdot K \}
\end{equation}

\noindent
\textbf{Schema Filling.} In multi-table schema retrieval, user queries often involve implicit joins which, although may sometimes be inferred by LLMs, are more often easily overlooked. For example, the query "How many schools with an average score in Math under 400 in the SAT test are exclusively virtual?" implicitly requires a join condition between \textit{schools.CDSCode} and \textit{satscores.cds} so that the relevant column can be correctly retrieved for subsequent reasoning. To address this, our retrieval framework explicitly incorporates relational keys. For the tables corresponding to the recalled schema set \( S \), we augment \( S \) with each table’s primary key (\( PK_i \)) and the pre-computed foreign keys (\( FK_i \)) that establish the shortest-path relationships among them. This ensures that essential join keys are included for downstream tasks.
\begin{equation}
S_{\text{}} = S \cup \bigcup_{T_i \in \mathcal{T}(S)} \big( \mathrm{PK}_i \cup \mathrm{FK}_i \big)
\end{equation}

\subsection{Cell Retrieval}

The second stage of FGTR is cell retrieval, which builds upon the schemas identified in the first stage. Rather than relying on structural understanding, this stage leverages the reasoning capabilities of large models to determine the required value ranges for each retrieved schema and extract the corresponding cells, as illustrated in the cell retrieval stage of Figure~\ref{fig:struct}. Finally, the relevant row segments are merged to construct the fine-grained sub-table necessary for answering the query. This process is carried out in three well-defined steps.

\noindent
\textbf{Range Parsing.}
The goal of range parsing is to determine the value scope of each required schema based on the results from the schema retrieval stage and the input query \( Q \), thereby facilitating the subsequent cell retrieval step. Schemas are categorized into two types: constrained columns and dependent columns. Constrained columns are associated with explicit values or conditions, and their value ranges are narrowed by parsing constraints embedded in the query, enabling more precise and efficient retrieval. In contrast, dependent columns do not contain specific constraints; their value ranges are assumed to include all possible values by default. Thus, additional filtering can be omitted during cell retrieval. This categorization significantly optimizes the retrieval execution process.

\noindent
\textbf{Cell Mapping.}
Cell mapping aims to identify table columns that satisfy specified constraint conditions. Although such constraints are typically provided in the query, direct value-based matching often fails to meet practical requirements. In real-world query scenarios, issues such as typographical errors, variations in linguistic expression, or users’ incomplete understanding of the database storage schema frequently occur, making exact string matching insufficient for robust mapping.
To address this, we propose an adaptive mapping mechanism:
For non-numeric constraints, we employ semantic similarity matching. Unique cell values are pre-encoded into vector representations. Given \(n\) constraint keywords derived from the query, we perform \(n\) parallel approximate nearest neighbor searches using HNSW \citep{malkov2018efficient}. We retrieve at least top-\(n\) candidates and include any additional cells whose similarity scores exceed a predefined threshold, ensuring robustness against lexical ambiguity.
For numeric constrained columns, we directly apply exact value matching, as numeric constraints typically exhibit lower ambiguity and benefit more from deterministic filtering, thereby improving computational efficiency.

\noindent
\textbf{Cell Merging.}
After obtaining the complete set of required columns \( S \) and the corresponding row ranges for each constrained column, we proceed to construct the query-relevant sub-table \( T_i' \). Specifically, for each constrained column \( c \in S_c \subseteq S \), a set of relevant row indices \( R_c \) is determined from the cell mapping results. For non-constrained (dependent) columns, the row scope is considered to be the full range of the table.

To form the final sub-table, we unify the row selections from all constrained columns. A row is included if it satisfies the constraints for at least one of the constrained columns. Therefore, the final set of row indices \( \mathcal{R}_i \) is the union of all individual row sets \( R_c \). The final sub-table \( T_i' \) is then constructed by selecting rows from \( T_i \) indexed by \( \mathcal{R}_i \) and restricting the columns to \( S \):
\begin{equation}
    T_i' = T_i \big[\, \mathcal{R}_i,\, S \,\big],
    \quad \text{where} \quad
    \mathcal{R}_i = \bigcup_{c \in S_c} R_c.
\end{equation}

\section{Experiments}
In this section, we evaluate both the baseline methods and our proposed FGTR framework on multi-table retrieval datasets as well as table-based downstream task datasets.
The objective of this evaluation is to address the following two research questions:
(1) To what extent does our approach—designed to mimic human reading behavior when interpreting tables—improve the performance of relevant sub-table retrieval?
(2) To what extent does effective relevant-table retrieval contribute to enhancing the performance of downstream table reasoning tasks?

\begin{table*}[t]
\centering
\small
\setlength{\tabcolsep}{10pt}
\caption{Schema retrieval results on Spider and BIRD datasets under different backbone models.
R, P, and $F_{2}$ denote recall, precision, and $F_{2}$, respectively.
* indicates results reported in the original papers.
\textdagger  indicates results from our reproduction.
\textit{@n} indicates voting with n predictions. \textit{+key} indicates schema filling is used.}
\label{tab:schema_retrieval}
  \begin{tabular}{lcccccccc}
\toprule
  \multirow{2}{*}{\textbf{Method}} & \multicolumn{4}{c}{\textbf{Spider}} & \multicolumn{4}{c}{\textbf{BIRD}} \\
  \cmidrule(lr){2-5} \cmidrule(lr){6-9}
  & \textbf{R} & \textbf{P} & \textbf{$F_{2}$} & \textbf{SR} & \textbf{R} & \textbf{P} & \textbf{$F_{2}$} & \textbf{SR} \\
\midrule
  Full-schema & 100 & 11.94 & 100 & 100 & 100 & 8.87 & 100 & 100 \\
  Gold-Based & 100 & 100 & 100 & 100 & 100 & 100 & 100 & 100 \\
\midrule
  \multicolumn{9}{l}{\textbf{\textit{Backbone (Llama3-8b)}}} \\
  C3 & 70.74 & 74.83 & 71.52 & 32.45 & 52.00 & 75.00 & 55.40 & 28.34 \\
  FGTR & 74.12 & 80.32 & 75.28 & 39.54 & 72.28 & 76.04 & 73.00 & 35.19 \\
  FGTR\textit{@3} & 76.25 & \underline{81.96} & 77.33 & 43.66 & 73.55 & 71.97 & 73.23 & 38.79 \\
  FGTR\textit{@5} & 83.23 & 71.44 & 80.57 & 49.57 & 77.52 & 73.18 & 76.61 & 46.33 \\
  FGTR\textit{@5+key} & 89.34 & 64.58 & 82.98 & 70.74 & 85.29 & \underline{78.82} & 82.19 & 70.64 \\
\midrule
  \multicolumn{9}{l}{\textbf{\textit{Backbone (Llama3-8b-FT)}}} \\
  FGTR@5+key & 91.52	&72.83	&87.05	&80.23 & 91.47 & 63.94 & 84.25 & 76.82 \\
\midrule
  \multicolumn{9}{l}{\textbf{\textit{Backbone (gpt-4o)}}} \\
  C3 & 91.39 & \textbf{87.82} & \underline{90.65} & 73.74 & 89.42 & \textbf{85.12} & \underline{88.53} & 69.32 \\
  CHESS & 93.82 & 72.86 & 88.72 & 76.29 & 91.49 & 67.56 & 85.45 & 78.26 \\
  RSL-SQL* & -&	-&	-&	- & \textbf{97.69}	& -	& -	& \textbf{94.32} \\
  RSL-SQL\textdagger & \underline{96.59} & 50.31 & 78.58 & \underline{88.44} & 97.13 & 35.01 & 72.00 & 89.54 \\
  TableRAG* & - & - & - & - & 95.30	&31.1	&67.50 & -\\
  TableRAG\textdagger & 92.48 & 31.14 & 66.39 & 85.62 & 90.20 & 44.49 & 56.91 & 79.32 \\
  FGTR\textit{@5+key} & \textbf{98.32} & 70.72 & \textbf{91.20} & \textbf{89.32} & \underline{97.27} & 72.00 & \textbf{90.89} & \underline{90.73} \\
\bottomrule
\end{tabular}
\end{table*}

\subsection{Experiment Setup}

\noindent
\textbf{Table Retrieval Dataset.}
Existing research on relevant sub-table retrieval lacks dedicated datasets that can directly evaluate retrieval performance. Moreover, most existing tabular task dataset consist of small tables with only 4-20 columns, making it difficult to accurately assess model performance on multi-table and medium-to-large-scale tabular structures.
To address this limitation, we first construct two benchmarks, SpiderQA and BirdQA, for the fine-grained retrieval task by utilizing the SPIDER\citep{yu2018spider} and BIRD\citep{li2024can} datasets. By parsing the gold SQL queries from these datasets, we extract the ground-truth columns and cells relevant to each question, which serve as the evaluation standard for our model’s retrieval performance. The detailed construction process is described in Appendix~\ref{app:dataset}.
For evaluating downstream performance, we use these two newly constructed datasets along with two widely-used tabular reasoning benchmarks: WikiTableQuestions (WTQ)\citep{pasupat2015compositional} for question answering and TabFACT\citep{chen2019tabfact} for fact verification.

\noindent
\textbf{Evaluation Metrics.}
On the SpiderQA and BirdQA datasets, we evaluate the quality of the retrieved sub-tables at both the schema and cell levels using \textbf{Precision}, \textbf{Recall}, \textbf{$\boldsymbol{F}_{2}$}, and \textbf{Strict Recall (SR)}. Strict Recall is a binary metric that scores 1 for complete recall of all ground-truth items and 0 otherwise. We use the $F_{2}$ (\(\beta=2\)) as our primary metric to emphasize recall. This decision is motivated by the fact that in retrieval-augmented systems, failing to retrieve relevant information leads to irrecoverable errors in downstream reasoning, whereas the inclusion of some irrelevant context is often less harmful.

For the end-to-end evaluation on SpiderQA, BirdQA, WikiTableQuestions, and TabFACT, we adopt standard task-specific metrics. We report EM accuracy for the question-answering tasks and standard accuracy for the fact verification task (TabFACT).

\noindent
\textbf{Baselines.}
We compare FGTR against state-of-the-art (SOTA) methods, which we categorize based on their retrieval granularity: schema-level and cell-level.

\textit{Schema-Level Retrieval.} This category includes methods that primarily function as schema-linking components in Text-to-SQL pipelines. They identify relevant columns by matching the query against table schemas. We select several strong baselines from this category to evaluate the schema retrieval stage of FGTR, including C3\citep{dong2023c3}, CHESS\citep{talaei2024chess}, RSL\citep{cao2024rsl}, and TableRAG\citep{chen2024tablerag}.

\textit{Cell-Level Retrieval.} This category encompasses methods that retrieve a fine-grained sub-table containing relevant rows and columns. While several methods exist, such as ITR and Tap4LLM, most are designed exclusively for single-table scenarios. As our work targets the more challenging and realistic multi-table retrieval setting, these single-table methods are not directly comparable. Therefore, we select TableRAG\citep{chen2024tablerag} as our primary baseline for cell-level retrieval. While many methods are strictly confined to single-table operations, the design of TableRAG allows it to be extended to the multi-table setting. We have adapted it accordingly for our experiments to ensure a fair and challenging SOTA comparison.

\subsection{Main Results}

We evaluate FGTR and several baseline methods on both the BIRD and Spider datasets for schema retrieval and cell retrieval performance.

\noindent
\textbf{Schema Retrieval.}
As presented in Table~\ref{tab:schema_retrieval}, FGTR consistently outperforms baselines across different backbones. Powered by GPT-4o, FGTR achieves a superior trade-off between recall and precision compared to schema-linking methods like RSL-SQL and C3. While baselines may achieve high recall, they often suffer from lower precision due to aggressive retrieval. FGTR, conversely, maintains state-of-the-art recall while significantly boosting precision (e.g., significantly higher $F_2$ on BIRD). This indicates that our strategy effectively filters irrelevant columns, providing a less redundant set of schemas for subsequent stages.

A key insight from our results is the relationship between Recall and Strict Recall (SR). SR, a challenging metric that requires the retrieval of the complete set of ground-truth schemas, is highly sensitive to any recall imperfections. We observe that when recall drops below 90

Furthermore, experiments with Llama3-8B demonstrate FGTR's adaptability to smaller open-source models. While the base model already performs competitively, fine-tuning via knowledge distillation yields substantial improvements. On the Spider dataset, this process boosts recall by 7.77 points (from 83.23

\noindent
\textbf{Cell Retrieval.} Following the schema retrieval stage, we now evaluate the performance of cell retrieval. This critical final step is responsible for filtering the data to its most relevant subset of rows based on the query's constraints. The effectiveness of this stage directly determines the quality of the final sub-table provided to the downstream task.

Table~\ref{tab:cell_retrieval} highlights the efficiency of our hierarchical approach. Existing methods like TableRAG operate on full-table encoding, which forces a trade-off: to maintain recall, they must retrieve a large volume of data, resulting in extremely low precision. In contrast, FGTR's adaptive mechanism leverages the constraints identified in the schema stage to precisely locate values. This allows FGTR to achieve a substantial improvement in precision compared to TableRAG, while retaining near-perfect recall.

This performance advantage is theoretically supported by computational complexity. By decomposing the task, FGTR reduces the retrieval complexity from $O(NM)$ (full table scan) to $O(N+M)$ (schema scan + targeted row scan). This efficiency enables the use of complex reasoning prompts that would be computationally prohibitive if applied to the entire table.

Overall, by successfully decomposing the complex retrieval task, FGTR's hierarchical retrieval strategy enables it to achieve high recall and high precision simultaneously. This demonstrates a strong comprehensive performance and establishes a new SOTA level across benchmarks by overcoming the noise inherent in single-stage methods.

\begin{table}[h!]
\centering
\caption{Cell retrieval results on Spider and BIRD datasets.  k indicates retrieval of the top-k most similar values, while \textit{adapt} refers to our adaptive cell mapping mechanism.}
\label{tab:cell_retrieval}
\setlength{\tabcolsep}{0pt}
\begin{tabular*}{\columnwidth}{@{\extracolsep{\fill}}lcccccccc}
\toprule
\multirow{2}{*}{\textbf{Method}} & \multicolumn{4}{c}{\textbf{Spider}} & \multicolumn{4}{c}{\textbf{BIRD}} \\
\cmidrule(lr){2-5} \cmidrule(lr){6-9}
 & \textbf{R} & \textbf{P} & \textbf{$F_{2}$} & \textbf{SR} & \textbf{R} & \textbf{P} & \textbf{$F_{2}$} & \textbf{SR} \\
\midrule
Full BD     & 100   & 8.07  & -     & 100  & 100   & 2.94  & -     & 100  \\
Gold-based  & 100   & 100   & 100   & 100  & 100   & 100   & 100   & 100  \\
\midrule
TableRAG    & 92.84 & 8.82  & 31.96 & 63.08 & 87.40 & 5.70  & 22.60 & 42.97 \\
FGTR\textit{(k=3)}      & 94.98 & \underline{32.74} & 68.82 & 72.73 & \underline{96.74} & \underline{26.17} & \underline{62.85} & \underline{76.13} \\
FGTR\textit{(k=5)}      & \underline{97.73} & 24.31 & \underline{69.93} & \underline{84.43} & \textbf{97.42} & 18.19 & 52.06 & \textbf{83.20} \\
FGTR\textit{(adapt)}      & \textbf{97.78} & \textbf{64.75} & \textbf{88.73} & \textbf{84.67} & 92.48 & \textbf{61.54} & \textbf{84.03} & 72.26 \\
\bottomrule
\end{tabular*}
\end{table}

\subsection{Downstream Task Experiment}
In this section, we investigate the impact of fine-grained table retrieval on the performance of downstream tasks, as well as how different retrieval scopes affect model effectiveness. We evaluate our approach on three table-based QA datasets and one table-based fact verification dataset, and report the QA accuracy of various retrieval models and FGTR under different retrieval scopes in Table~\ref{tab:downstream}.
From Table~\ref{tab:downstream}, we observe that both schema retrieval and cell retrieval lead to consistent improvements in downstream task performance compared to using the entire table without retrieval. Overall, the improvement brought by schema retrieval is more pronounced. This may be attributed to the lower accuracy of range parsing when schema retrieval is omitted, indirectly confirming the importance of performing schema retrieval as the first-stage operation.

On the other hand, cell retrieval plays a crucial role in reducing token consumption and eliminating redundant information. Our analysis shows that only after applying cell retrieval can all four datasets be fully accommodated within the context window of LLMs, thereby enabling end-to-end QA capability.

Furthermore, we aim to examine whether better retrieval performance translates into better end-to-end results in downstream tasks. To this end, we compare the downstream QA accuracy obtained from the table retrieval results of TableRAG, Dater, and FGTR. Across all four tasks, FGTR achieves the highest EM accuracy, demonstrating that our model’s retrieval process not only enhances retrieval quality but also yields tangible improvements in the end-to-end performance of table-based downstream tasks.

\begin{table}[t]
\centering
\small
\setlength{\tabcolsep}{0pt}
\newcommand{\cmark}{\ding{51}}
\newcommand{\xmark}{\ding{55}}
\caption{Ablation study of the schema retrieval (SR) and cell retrieval (CR) components. Using both components in our FGTR model yields the best performance across all four downstream QA and fact-verification datasets. The performance of two baseline methods, TableRAG and Dater, is included for comparison. The "-" symbol indicates that the input table was too large to fit into the model's context window.}
\label{tab:downstream}
\begin{tabular*}{\columnwidth}{@{\extracolsep{\fill}}l cc cccc}
\toprule
\multirow{2}{*}{\textbf{Method}} & \multicolumn{2}{c}{\textbf{Components}} & \multicolumn{4}{c}{\textbf{Downstream Accuracy (\%)}} \\
\cmidrule(lr){2-3} \cmidrule(lr){4-7}
& \textbf{SR} & \textbf{CR} & \textbf{BirdQA} & \textbf{SpiderQA} & \textbf{TabFACT} & \textbf{WTQ} \\
\midrule
\multirow{4}{*}{FGTR}
& \cmark & \cmark & \textbf{53.76} & \textbf{78.26} & \textbf{76.25} & \textbf{60.27} \\
& \cmark & \xmark & --    & 75.21 & 74.84 & 55.47 \\
& \xmark & \cmark & 46.72 & 74.80 & 67.20 & 56.28 \\
& \xmark & \xmark & --  & 66.08 & 62.90 & 53.00 \\
\midrule
TableRAG & \multicolumn{2}{c}{--} & 45.50 & 48.61 & 69.42 & 56.63 \\
Dater    & \multicolumn{2}{c}{--} & -- & -- & 70.3	&46.9\\
\bottomrule
\end{tabular*}
\end{table}

\begin{table*}[t!]
\centering
\small
\caption{Ablation study of FGTR on Spider and BIRD datasets for both schema retrieval and cell retrieval.}
\label{tab:ablation}
\begin{tabular}{lcccccccccccc}
\toprule
\multirow{2}{*}{\textbf{Method}}
& \multicolumn{6}{c}{\textbf{Spider}}
& \multicolumn{6}{c}{\textbf{BIRD}} \\
\cmidrule(lr){2-7} \cmidrule(lr){8-13}
& \multicolumn{3}{c}{\textbf{Schema Retrieval}} & \multicolumn{3}{c}{\textbf{Cells Retrieval}}
& \multicolumn{3}{c}{\textbf{Schema Retrieval}} & \multicolumn{3}{c}{\textbf{Cells Retrieval}} \\
\cmidrule(lr){2-4} \cmidrule(lr){5-7} \cmidrule(lr){8-10} \cmidrule(lr){11-13}
  & \textbf{R} & \textbf{P} & \textbf{$F_{2}$} & \textbf{R} & \textbf{P} & \textbf{$F_{2}$} & \textbf{R} & \textbf{P} & \textbf{$F_{2}$} & \textbf{R} & \textbf{P} & \textbf{$F_{2}$} \\
\midrule
FGTR & \textbf{98.32} & 70.72 & \textbf{91.20} & \textbf{97.78} & \textbf{84.75} & \textbf{94.86} & \textbf{97.27} & 72.00 & \textbf{90.89} & \textbf{92.48} & \textbf{90.54} & \textbf{92.09} \\
\midrule
-w/o column explaining & 96.63 & 68.41 & 89.27 & - & - & - & 92.17 & 70.30 & 86.84 & - & - & - \\
-w/o question parsing & 95.71 & 59.73 & 85.42 & - & - & - & 93.54 & 65.20 & 86.06 & - & - & - \\
-w/o schema mapping & 91.62 & 76.38 & 88.21 & - & - & - & 89.08 & 81.33 & 87.43 & - & - & - \\
-w/o schema filling & 93.52 & \textbf{79.18} & 90.20 & - & - & - & 80.74 & \textbf{85.56} & 80.43 & - & - & - \\
-w/o range parsing & - & - & - & 81.33 & 31.38 & 61.67 & - & - & - & 71.66 & 10.12 & 32.30 \\
\bottomrule
\end{tabular}
\end{table*}

\subsection{Ablation Experiment Results}

To investigate the contribution of each technical component in FGTR, we conduct a series of ablation studies as summarized in Table~\ref{tab:ablation}.

\textbf{w/o column explaining}, where the model relies solely on original schema names, we observe a recall drop of 1.7 points on Spider and 5 points on BIRD. This decline attributes to the prevalence of abbreviated column names in BIRD, verifying that column explaining is critical for resolving semantic ambiguity. Similarly, \textbf{w/o question parsing}, bypassing the decomposition step leads to a substantial decrease in precision. This degradation suggests that question parsing plays an essential role in filtering noise and guiding the model toward precise schema-level targets.

Next, \textbf{w/o schema mapping}, We constrain the model to execute a single retrieval request. The results indicate a significant drop in recall. The observed slight improvement in precision primarily stems from the incomplete retrieval coverage, and secondly, \textbf{w/o schema filling}, the results indicate that precision improves slightly, but recall drops significantly—especially on the BIRD dataset. This suggests that while schema filling may introduce some redundancy that slightly reduces precision, it is highly beneficial for accurately retrieving schemas involving multi-table joins. Notably, the overall $F_{2}$ improves by 3 points on the Spider dataset and 10 points on the BIRD dataset, confirming the necessity and effectiveness of schema filling for complex, multi-table database retrieval scenarios.

Finally, in the cell retrieval stage, \textbf{w/o range parsing} causes substantial drops in both recall and precision. This occurs because the model generates queries without specific column constraints, matching them against all schema columns and introducing significant noise. The results demonstrate that range parsing is effective in determining precise retrieval scopes, thereby enhancing both efficiency and accuracy.

\subsection{Performance Analysis Across Table Scales}
\begin{figure}[h]
  \centering
  \includegraphics[width=\columnwidth]{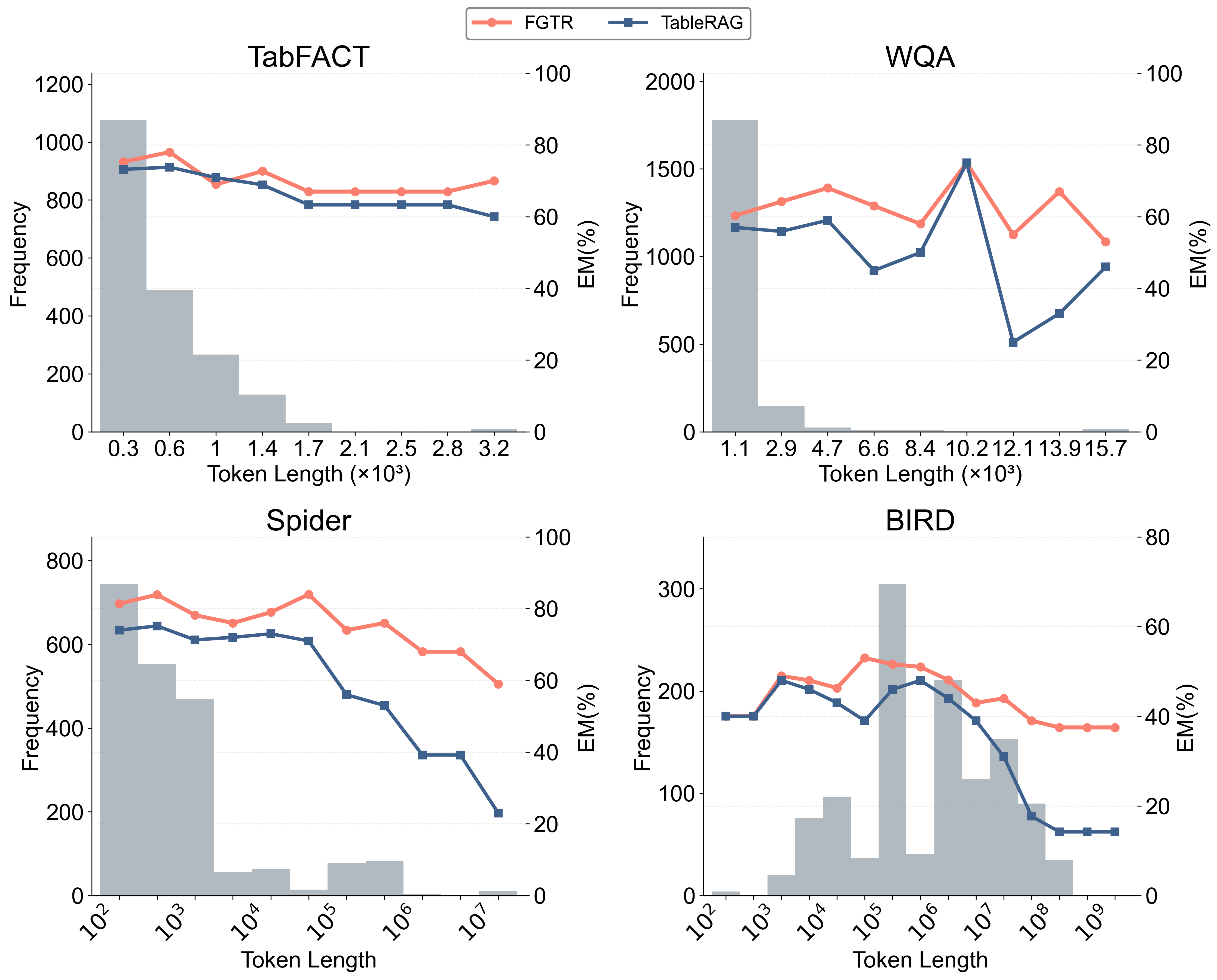}
  \caption{Performance comparison of FGTR and TableRAG across different table sizes. The x-axis represents table size (number of tokens). The left y-axis and the bar chart show the distribution of table sizes in our datasets. The right y-axis corresponds to the line plots, which compare the EM accuracy of FGTR against TableRAG on four downstream tasks. Each subplot corresponds to a different dataset.}
  \label{fig:combined_distribution}
\end{figure}
To examine how table size affects model performance, we evaluate FGTR and several baseline methods on four downstream table-related datasets across varying table sizes. Specifically, we measure the EM accuracy under different token-scale settings to analyze the robustness of fine-grained table retrieval approaches as table size increases.
The results, shown in Figure~\ref{fig:combined_distribution}, indicate that FGTR and TableRAG perform comparably when table sizes are relatively small. However, as the table size grows---particularly in the Spider and BIRD datasets where the number of tokens exceeds $10^7$---TableRAG exhibits a sharp performance degradation, while FGTR maintains significantly higher stability.
This demonstrates FGTR's superior resilience to large-scale tables: its two-step reasoning design allows the model to efficiently focus on relevant schema and cell subsets rather than encoding the entire table. In contrast, methods like TableRAG, which rely on full-table encoding, suffer from increased computational complexity and reduced retrieval precision as table size expands.
Overall, these results highlight the scalability and efficiency advantages of FGTR, demonstrating its strong robustness in handling relational tables of varying sizes and complexity.

\section{Conclusion}
In this paper, we introduced FGTR, a novel method that addresses the challenges of fine-grained table retrieval by employing a hierarchical, human-like reasoning strategy. By decomposing the complex task into two manageable stages—schema retrieval followed by cell retrieval—FGTR effectively leverages the reasoning capabilities of LLMs to navigate structured data. Our extensive experiments on the newly constructed SpiderQA and BirdQA benchmarks demonstrate that this decompositional approach is highly effective. FGTR significantly outperforms state-of-the-art baselines by achieving a superior balance between high recall and high precision, leading to the retrieval of more concise and relevant sub-tables. This work validates that guiding LLMs through a hierarchical reasoning process is a more potential strategy than relying on single-stage, full-table encoding, paving the way for more accurate and efficient table-based reasoning systems.

\bibliographystyle{ACM-Reference-Format}
\bibliography{references}

\appendix

\section{Case Study}\label{app:case}

\begin{figure}[h]
  \centering
  \includegraphics[width=\columnwidth]{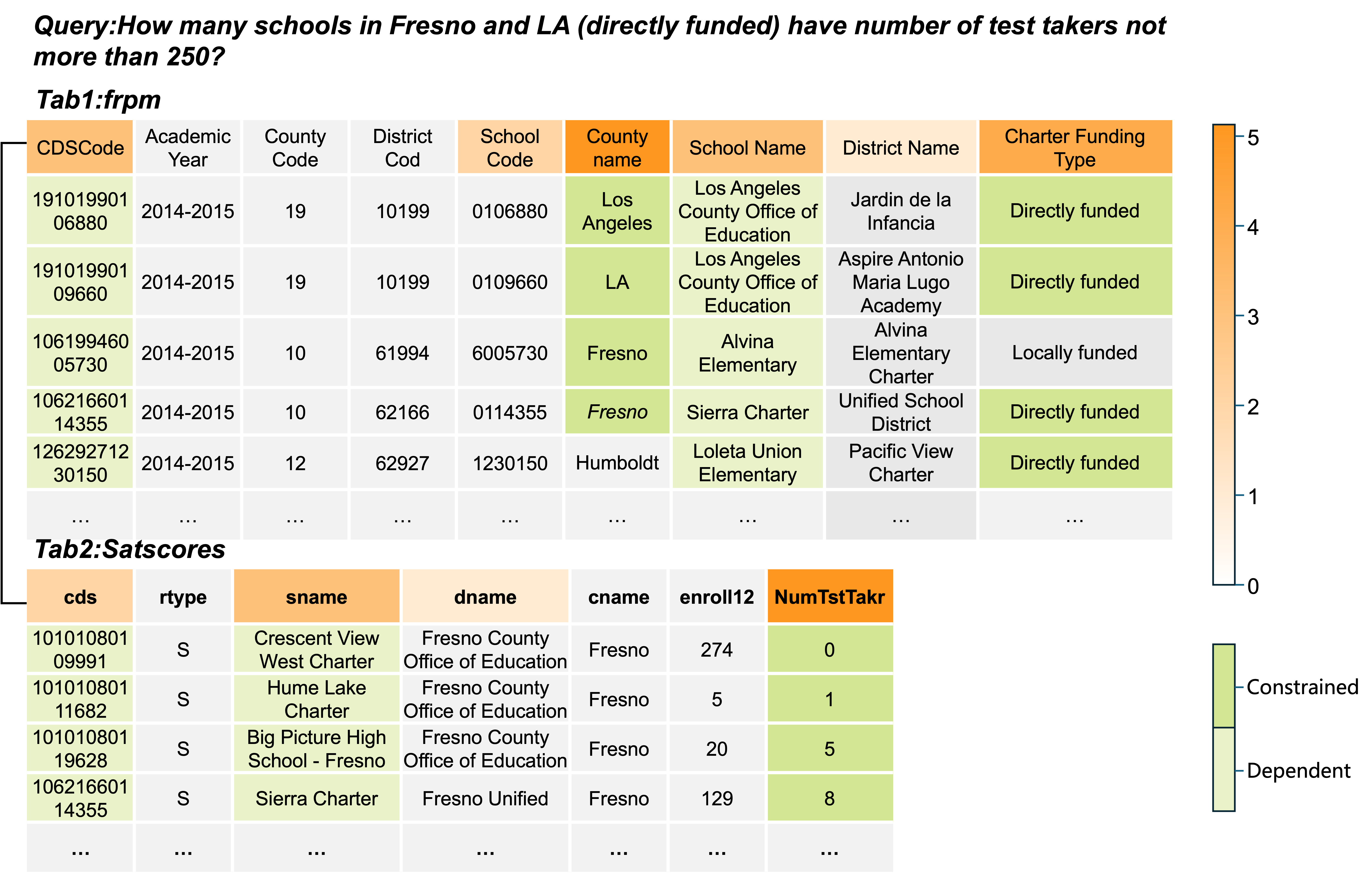}
  \caption{Oriainal tables from BirdQA with FGTR relecance heat-map. In the schema retrieval phase, deeper orange indicates higher frequency fields, while in the cell retrieval phase, green intensity distinguishes between constrained (dark) and dependent (light) cells.}
  \label{fig:case1}
\end{figure}

\begin{figure}[h]
  \centering
  \includegraphics[width=\columnwidth]{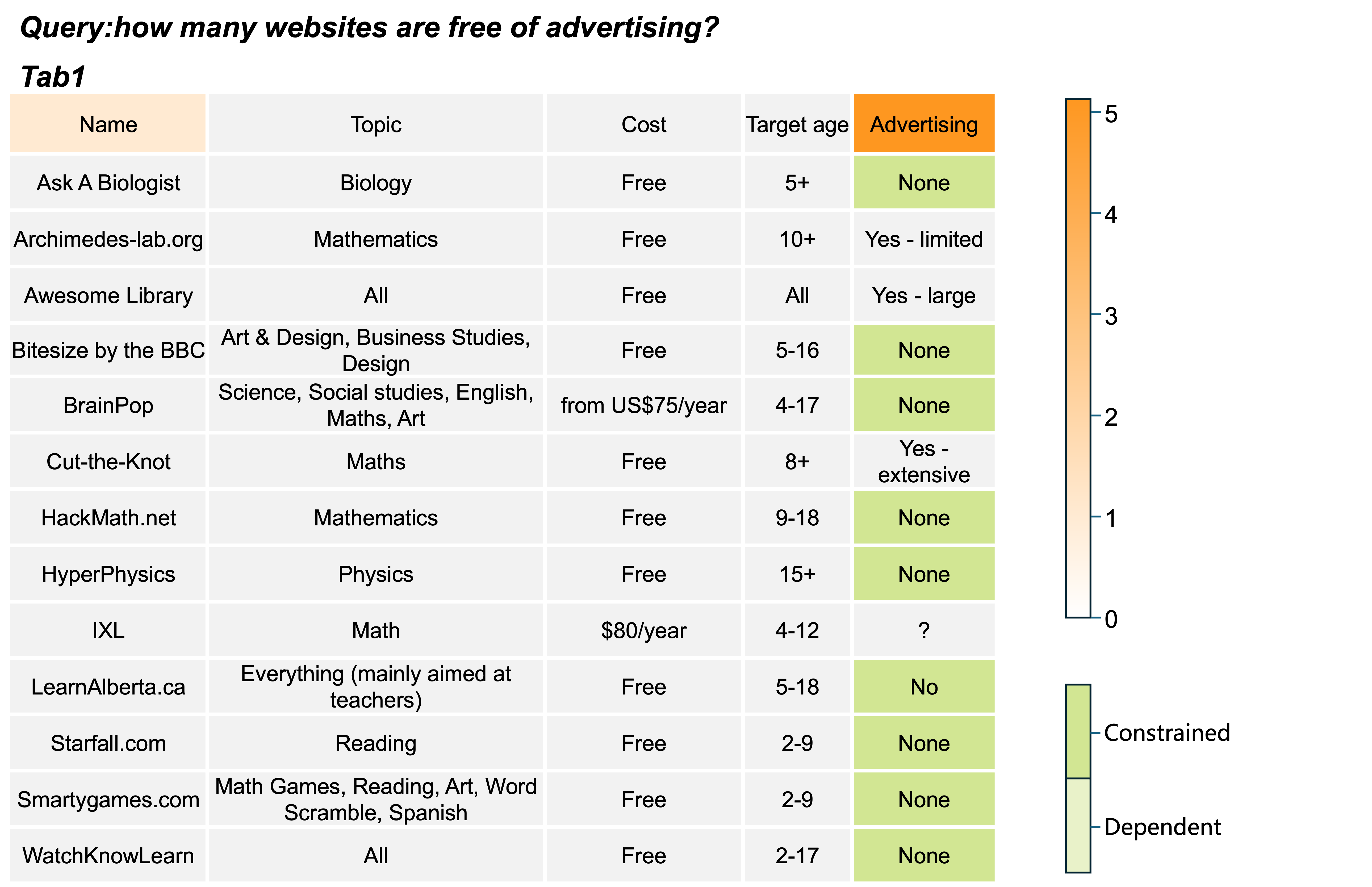}
  \caption{Original tables from WTQ with FGTR relecance heat-map.}
  \label{fig:case2}
\end{figure}

To strictly demonstrate how FGTR performs fine-grained information retrieval across multiple tables, we present two concrete case studies in Figure~\ref{fig:case1} and Figure~\ref{fig:case2}. The visualization employs dual-color heatmaps to trace the retrieval trajectory: in the \textit{Schema Mapping} phase, the saturation of orange correlates with the recall frequency of each field across $k=5$ iterations; in the \textit{Cell Retrieval} phase, green regions denote the final retrieval scope, identifying constrained fields (dark green) and dependent fields (light green).

Figure~\ref{fig:case1} illustrates a case from the BirdQA dataset, in the \textit{Schema Mapping} phase with a consistency threshold $\theta=0.6$, essential conditional fields such as \textit{Country Name}, \textit{Charter Funding Type}, and \textit{NumTstTakr} are consistently recalled (high saturation).
Conversely, sporadic noise like \textit{School Code} and \textit{District Name} is effectively filtered out.
It is worth noting that although \textit{School Name} and \textit{sname} are strictly redundant for this query, their retrieval was primarily triggered by the keyword "schools". The presence of such redundancy does not affect the overall retrieval performance; on the contrary, ensuring the complete recall of correct content is of greater importance.
Furthermore, regarding table connectivity, the critical join key \textit{cds}—which was initially missed in the mapping phase—was successfully retrieved in the subsequent \textit{Schema Filling} phase.
In this case, FGTR reduces data redundancy by 98

Figure~\ref{fig:case2} illustrates a case from the WTQ dataset, where the query asks for the count of websites with free advertising. Given the table metadata, FGTR accurately identifies that only the \textit{Advertising} column is necessary to derive the answer. Crucially, through robust semantic matching, the model correctly pinpoints cells containing "None" and "No" as satisfying the "free" condition, thereby retrieving all—and only—the content required for the answer. In the subsequent QA test, feeding the raw source table to the LLM resulted in an erroneous answer of "8". In contrast, providing the FGTR-retrieved sub-table yielded the correct answer of "9". This demonstrates that by supplying the LLM with a concise, noise-free representation of the data, FGTR significantly reduces cognitive interference and enhances reasoning accuracy.

\section{Dataset Construction and Augmentation}\label{app:dataset}

We present our methodology for constructing and augmenting the benchmark datasets, with detailed statistics summarized in Table~\ref{tab:dataset_statistics_detailed}.

\begin{table}[t]
  \centering
  \caption{
    Statistical overview of the benchmark datasets.
    The "Overall" columns describe the general and structural properties of the source databases, including total counts of databases and questions, the average number of tables per database, and the average columns and rows per table.
    The "Per Question" columns detail the complexity of the ground-truth schema for an average question, reporting the mean number of relevant tables and columns.
}
\label{tab:dataset_statistics_detailed}
  \resizebox{\columnwidth}{!}{
  \begin{tabular}{l rrrrr rr}
  \toprule
  \multirow{2}{*}{\textbf{Dataset}} & \multicolumn{5}{c}{\textbf{Overall}} & \multicolumn{2}{c}{\textbf{Per Question}} \\
  \cmidrule(lr){2-6} \cmidrule(lr){7-8}
  & \textbf{\# DBs} & \textbf{\# Questions} & \textbf{Tables/DB} & \textbf{Cols/Table} & \textbf{Rows/Table} & \textbf{Tables/Q} & \textbf{Cols/Q} \\
  \midrule
  SpiderQA & 166 & 965 & 5.13 & 5.01 & 17 & 1.52 & 2.99 \\
  BIRDQA   & 95 & 1,182 & 7.25 & 10.66 & 55,448 & 2.17 & 4.95 \\
  WTQ      & 1,876 & 2,000 & 1.0 & 6.41 & 28.01 & 1.00 & - \\
  TabFACT  & 1,695 & 2,000 & 1.0 & 6.25 & 13.43 & 1.00 & - \\
  \bottomrule
  \end{tabular}
  }
\end{table}

\noindent
\textbf{Dataset Construction.} A primary challenge in evaluating fine-grained table retrieval is the lack of dedicated benchmarks. To address this, we construct SpiderQA and BIRDQA, two new large-scale, cross-domain datasets derived from the widely-adopted Text-to-SQL benchmarks, Spider and BIRD, respectively. These datasets are designed to mirror real-world scenarios and consist of a question, its corresponding database, and a gold SQL query.

Our core contribution lies in creating a ground-truth sub-table for each question, which serves as the gold standard for fine-grained retrieval. The construction process is as follows: for each gold SQL query, we syntactically parse it to extract all unique columns, which are stored in a Table.Column format and serve as the ground truth for the schema retrieval stage. We then formulate a new SQL query by replacing the original SELECT clause with the full set of extracted columns, while keeping all other clauses (e.g., FROM, WHERE) unchanged. The execution of this new query yields a precise, minimal sub-table containing all and only the information required to answer the original question. This resulting sub-table becomes the ground-truth for evaluating retrieval performance.

Our final datasets include only those samples where the sub-table generation is successful and the resulting data is amenable to question-answering tasks. Furthermore, by running the original standard SQL query, we obtain the final answer to the question. This transforms SpiderQA and BIRDQA into fully-fledged table question-answering datasets, enabling the evaluation of not only the retrieval stage but also the performance of downstream, end-to-end tasks. To complement our evaluation, we also include the established WTQ and TabFACT benchmarks for a comprehensive assessment of downstream reasoning capabilities.

\noindent
\textbf{Dataset Augmentation.}
To simulate real-world scenarios where data exhibits lexical ambiguity and to rigorously evaluate the model's robustness in fuzzy semantic matching, we perform a targeted data augmentation. This strategy involves applying synonym replacement to the textual content within the database tables. Specifically, for entities relevant to a given question, we programmatically substitute their values in the database with common synonyms, acronyms, or abbreviations.

For instance, a cell value such as "Los Angeles" in a "country" column might be replaced with its common abbreviation "LA". This process intentionally introduces a semantic gap between the phrasing in the user's question and the stored values in the table. By doing so, we create a more challenging benchmark to assess the model's ability to handle semantic variations and its resilience to lexical diversity during retrieval.

\section{Details of Prompt}\label{app:prompt}
In each prompt of our method, we explicitly require the model to generate a reasoning explanation alongside its output. This mechanism, akin to Chain-of-Thought (CoT) prompting~\citep{wei2022chain}, not only enhances interpretability and stability by exposing the mapping rationale for debugging but also provides high-quality supervision data for fine-tuning smaller models.Due to space limitations, we present a representative prompt example here.

\begin{tcolorbox}[
  colback=gray!5!white,
  colframe=gray!75!black,
  title=Schema Mapping Prompt,
  enhanced,
  breakable,
  sharp corners,
  listing only,
  listing options={
    basicstyle=\ttfamily\small,
    breaklines=true,
    breakatwhitespace=false,
    columns=fullflexible,
    keepspaces=true,
    showstringspaces=false,
    escapeinside={(*@}{@*)}
  }
]
\# You are an expert on data analysis. Your task is to examine the provided database schema, understand the posed question, and use column examples to pinpoint the specific columns within tables that are essential to solve the problem.

\#\# JSON format database schema, represented as column name: column description:
\{TABLESTRUCTURE\}

\#\# The column examples include entites that may be necessary to solve the problem, serving as a key hint for retrieving the required columns. Please note:

-1. Please identify appropriate columns to cover all entities mentioned in the hints, Please ensure the selected columns strictly adhere to the given columns names.
-2. The entites in hint may not contain all the required columns.
-3. A single entites in the set may actually require multiple columns to be fully satisfied.

\#You need to not only select the required columns, but also explain in detail why each column is needed.

\#Your answer should strictly follow the following json format.

\{\{
"reasoning": "",
"columns": ["table\_name\_i.column\_name\_j", ...]
\}\}

\# Your Answer should just be a json:

\end{tcolorbox}

\typeout{get arXiv to do 4 passes: Label(s) may have changed. Rerun}
\end{document}